\documentclass[backref, breaklinks, twocolumn, colorlinks]{aastex631}   
\hypersetup{linkcolor=blue,citecolor=blue,filecolor=cyan,urlcolor=black}
 
\usepackage{latexsym,amsmath,amssymb}
\usepackage{verbatim}
\usepackage{multirow}
\usepackage{color}
\usepackage[hang]{footmisc}
\usepackage{soul,xcolor}
\usepackage{afterpage}
\usepackage{savesym}
\savesymbol{tablenum}
\usepackage{siunitx}
\restoresymbol{SIX}{tablenum}

\bibpunct{(}{)}{;}{a}{}{,}

\newcommand {\swift} {\textsl{Swift}}

\newcommand {\nustar} {\textsl{NuSTAR}}

\newcommand {\nicer} {\textsl{NICER}}
\newcommand {\ixpe} {\textsl{IXPE}}

\def \rsun {\ifmmode$R$_{\odot}\else R$_{\odot}$}

\def \hcm {\hbox {\ifmmode $ atoms cm$^{-2}\else atoms cm$^{-2}$\fi}}

\def\approxgt{\mathrel{\hbox{\rlap{\lower.55ex \hbox {$\sim$}}
        \kern-.3em \raise.4ex \hbox{$>$}}}}
\def\approxlt{\mathrel{\hbox{\rlap{\lower.55ex \hbox {$\sim$}}
        \kern-.3em \raise.4ex \hbox{$<$}}}}

\def \arcsec {\hbox{$^{\prime\prime}$}}

\newcommand\T{\rule{0pt}{2.6ex}}
\newcommand\B{\rule[-1.2ex]{0pt}{0pt}}

\def \xteeigthteen {XTE\,J1810$-$197}
\def \swifteigthteen {Swift\,J1818$-$1607}
\def \seventeenfourtyfive {SGR\,J1745$-$2900}
\def \sixteentwentytwo {PSR\,J1622$-$4950}

\def \src {1E~1841$-$045}
\def \sgrnineteen {SGR\,1935+2154}

\begin{document}
\setstcolor{red}

\title{\rm \uppercase{Timing and spectral evolution of the magnetar \src\ in outburst}}

%
\author[0000-0002-7991-028X]{George~Younes}
\affiliation{Center for Space Sciences and Technology, University of Maryland, Baltimore County, Baltimore, MD 21250}
\affiliation{Astrophysics Science Division, NASA Goddard Space Flight Center, Greenbelt, MD 20771, USA}
\affiliation{Center for Research and Exploration in Space Science and Technology, NASA Goddard Space Flight Center, Greenbelt, MD 20771, USA}

\author{Samuel~K~Lander}
\affiliation{School of Engineering, Mathematics and Physics, University of East Anglia, Norwich, NR4 7TJ, U.K.}

\author[0000-0003-4433-1365]{Matthew~G.~Baring}
\affiliation{Department of Physics and Astronomy - MS 108, Rice University, 6100 Main Street, Houston, Texas 77251-1892, USA}

\author[0009-0006-3567-981X]{Marlon L. Bause}
\affiliation{Max Planck Institut für Radioastronomie, Auf dem Hügel 69, 53121 Bonn, Germany}

\author[0000-0002-0254-5915]{Rachael~Stewart}
\affiliation{Department of Physics, The George Washington University, Washington, DC 20052, USA, gyounes@gwu.edu}

\author[0009-0008-6187-8753]{Zaven Arzoumanian}
\affiliation{Astrophysics Science Division, NASA Goddard Space Flight Center, Greenbelt, Maryland 20771, USA}

\author[0000-0001-9268-5577]{Hoa~Dinh Thi}
\affiliation{Department of Physics and Astronomy - MS 108, Rice University, 6100 Main Street, Houston, Texas 77251-1892, USA}

\author[0000-0003-1244-3100]{Teruaki Enoto}
\affiliation{Department of Physics, Kyoto University, Kitashirakawa Oiwake, Sakyo, Kyoto 606-8502, Japan}

\author[0000-0001-7115-2819]{Keith Gendreau}
\affiliation{Astrophysics Science Division, NASA Goddard Space Flight Center, Greenbelt, Maryland 20771, USA}

\author[0000-0002-3531-9842]{Tolga G\"uver}
\affiliation{Istanbul University, Science Faculty, Department of Astronomy and Space Sciences, Beyaz\i t, 34119, Istanbul, Turkey}
\affiliation{Istanbul University Observatory Research and Application Center, Istanbul University 34119, Istanbul Turkey}

\author[0000-0001-6119-859X]{Alice~K.~Harding}
\affiliation{Theoretical Division, Los Alamos National Laboratory, Los Alamos, NM 87545, USA}

\author[0000-0002-6089-6836]{Wynn C. G. Ho}
\affiliation{Department of Physics and Astronomy, Haverford College, 370 Lancaster Avenue, Haverford, PA 19041, USA}

\author[0000-0001-8551-2002]{Chin-Ping Hu}
\affiliation{Department of Physics, National Changhua University of Education, Changhua 50007, Taiwan}

\author{Alex~van~Kooten}
\affiliation{Department of Physics, The George Washington University, Washington, DC 20052, USA, gyounes@gwu.edu}

\author[0000-0003-1443-593X]{Chryssa~Kouveliotou}
\affiliation{Department of Physics, The George Washington University, Washington, DC 20052, USA, gyounes@gwu.edu}
\affiliation{Astronomy, Physics and Statistics Institute of Sciences (APSIS), The George Washington University, Washington, DC 20052, USA}

\author[0000-0002-7574-1298]{Niccol\`o Di Lalla}
\affiliation{Department of Physics and Kavli Institute for Particle Astrophysics and Cosmology, Stanford University, Stanford, CA 94305, USA}
\author{Alexander~McEwen}
\affiliation{Center for Space Sciences and Technology, University of Maryland, Baltimore County, Baltimore, MD 21250}

\author[0000-0002-6548-5622]{Michela Negro}
\affiliation{Department of Physics \& Astronomy, Louisiana State University, Baton Rouge, LA 70803, USA}

\author[0000-0002-0940-6563]{Mason Ng}
\affiliation{Department of Physics, McGill University, 3600 rue University, Montr\'{e}al, QC H3A 2T8, Canada}
\affiliation{Trottier Space Institute, McGill University, 3550 rue University, Montr\'{e}al, QC H3A 2A7, Canada}

\author[0000-0001-7128-0802]{David M. Palmer}
\affiliation{Los Alamos National Laboratory, Los Alamos, NM 87544, USA}
\affiliation{New Mexico Consortium, Los Alamos, NM, 87544, USA}

\author[0000-0002-3775-8291]{Laura G. Spitler}
\affiliation{Max Planck Institut für Radioastronomie, Auf dem Hügel 69, 53121 Bonn, Germany}

\author[0000-0002-9249-0515]{Zorawar Wadiasingh}
\affiliation{Department of Astronomy, University of Maryland, College Park, Maryland 20742, USA}
\affiliation{Astrophysics Science Division, NASA Goddard Space Flight Center, Greenbelt, MD 20771, USA}
\affiliation{Center for Research and Exploration in Space Science and Technology, NASA Goddard Space Flight Center, Greenbelt, MD 20771, USA}

\begin{abstract}

We present the timing and spectral analyses of the \nicer, \nustar, and \ixpe\ observations of the magnetar \src\ covering 82~days following its August 2024 bursting activity as well as radio observations utilizing MeerKAT and Effelsberg. We supplement our study with a historical \nustar\ and all 2024 pre-outburst \nicer\ observations. The outburst is marked by an X-ray flux enhancement of a factor 1.6 compared to the historical level, predominantly driven by a newly-formed non-thermal emitting component with a photon index $\Gamma=1.5$. This flux showed a 20\% decay at the end of our monitoring campaign. The radio monitoring did not reveal any pulsed radio emission with an upper-limit of 20~mJy and 50~mJy~ms on the mean flux density and single pulse fluence, respectively. We detect a spin-up glitch at outburst onset with a $\Delta\nu=6.1\times10^{-8}$~Hz and a $\Delta\dot{\nu}=-1.4\times10^{-14}$~Hz~s$^{-1}$, consistent with the near-universality of this behavior among the continuously-monitored magnetars. Most intriguingly, the \src\ 2-10~keV pulse profile is markedly different compared to pre-outburst; it shows a new, narrow (0.1~cycles) peak that appears to shift towards merging with the main, persistently-present, pulse. This is the second case of pulse-peak migration observed in magnetars after SGR 1830$-$0645, and the two sources exhibit a similar rate of phase shift. This implies that this phenomenon is not unique and might present itself in the broader population. The newly-formed peak for \src\ is non-thermal, with emission extending to $\gtrsim20$~keV, in contrast to the case of SGR 1830$-$0645. Our results are consistent with an untwisting magnetic field bundle with migration towards the magnetic pole, perhaps accompanied by plastic motion of the crust.

\vphantom{)}
\vspace{70pt}
\end{abstract}

\section{Introduction}
\label{Intro}

Magnetars represent a subset of the isolated neutron star (INS) population, possessing dipole magnetic fields with strength reaching $10^{15}$~G \citep{kouveliotou98Nat:1806}. This large magnetic energy reservoir gives magnetars unique characteristics rarely present in other types of INSs, chief among them are their outburst episodes. The onset of such is commonly manifested by the emission of a few to several hundred bright ($E_{\rm X}\sim10^{37-41}$~erg), short (duration in the range 0.01-1~second), hard X-ray bursts \citep{collazzi15ApJS}. Concurrently, the magnetar soft and hard X-ray quiescent emission increases by as many as three orders of magnitude \citep{cotizelati18MNRAS}, accompanied by significantly higher surface temperatures, strong temporal variability in the form of timing noise, glitch activity, and altered pulse shape and fraction. This diverse phenomenology is typically interpreted in the context of the decay of strong crustal B-fields, causing magnetic stresses on the star surface that could exceed the yield strain of the solid crust \citep[e.g.,][]{lander15mnras}. This, in turn, can result in crustal deformation, thermal energy deposition leading to new hot spots on the star surface, and twisted external B-field loops \citep[see, e.g.,][]{thompson96ApJ:magnetar, harding99ApJ:mag, thompson02ApJ:magnetars, vigano13MNRAS, kaspi17:magnetars, younes22ApJ:ppm}.

\src\ is a persistently X-ray bright magnetar discovered with the Advanced Satellite for Cosmology and Astrophysics \citep{vasisht97ApJ}. Its timing characteristics imply a surface dipolar field of about $7\times10^{14}$~G, a spin-down age of about 4.6~kyr, and a rotational-energy loss $\sim10^{33}$~erg~s$^{-1}$. Its soft and hard X-ray luminosities exceed $10^{35}$~erg~s$^{-1}$ \citep[e.g.,][]{kuiper04ApJ:1841}, well in excess of its $|\dot{E}|$. It is located at the center of the X-ray and radio-bright supernova remnant (SNR) Kes 73, confirming its spin-down inferred young age. The \src\ broadband X-ray spectrum is well described with the canonical quasi-thermal soft X-ray and non-thermal hard X-ray tail model \citep{kuiper04ApJ:1841, an13ApJ:1841, an15ApJ:1841, 2017ApJS..231....8E}. 

Since its discovery, \src\ has displayed several episodes of magnetar bursting activity including the one that established it as part of the magnetar family \citep{gavriil02Natur:AXPs}, yet never was its persistent emission observed to be in an enhanced state \citep{kumar10ApJ:1841, lin11ApJ:1E1841, dib14ApJ}. On 2024 August 21, \src\ entered another active period which was first observed with Swift/BAT \citep{dichiara24ATel16784}, and later confirmed with Fermi/GBM and NICER \citep{roberts24ATel16786, ng24ATel16789}, among other high energy instruments. Follow-up \nicer, \nustar, and \swift\ observations revealed an enhanced 2-70 keV flux and a noticeably more complex soft and hard X-ray pulse profiles \citep{younes24ATel16802}, highlighting the first radiative outburst from the source in conjunction with its bursting emission.

In this paper, we report on the X-ray timing and spectral evolution of the magnetar \src\ during its 2024 August outburst through the analysis of \nicer, \nustar, and \ixpe\ observations covering the period 2024 August 21 to November 11. For completeness and comparison purposes, we also analyze all the \nicer\ observations conducted in 2024, and a historical \nustar\ observation taken in 2012. Lastly, we report on 2024 radio obserations of the source taken with MeerKAT and Effelsberg. Section~\ref{sec:obs} summarizes the observations and data reduction procedures. Section~\ref{sec:res} details our findings. Section~\ref{sec:disc} discusses our results in the context of surface and magnetospheric dynamics during magnetar outbursts.

\section{Observations, data reduction, and analysis procedures}
\label{sec:obs}

\subsection{X-rays}

We analyze the four Nuclear Spectroscopic Telescope ARray (\nustar;\, \citealt{harrison13ApJ:NuSTAR}) observations of \src\ that were taken post-outburst, which occurred over the course of 2.5 months from 2024 August 29, 8 days after the Swift-BAT announcement of the most recent activity from \src, until November 11. Each observation is roughly 50 ks in exposure. For comparison with the quiescent state, we also analyze the \nustar\ observation from 2012 November 9 (id 30001025002) which also had an exposure of about 50~ks (Table~\ref{tab:specres}). We perform the data reduction, calibration, and high-level product extraction, e.g., source and background cleaned event files, spectra, and light curves utilizing {\tt NuSTARDAS} software version 2.1.2 as part of HEASoft (version 6.33.1). We utilize a circular region with a radius of 45\arcsec\ centered on the central brightest pixel to extract source events, while background events were considered from a source-free circular region with a radius of 60\arcsec, on the same CCD (Charge Coupled Device) as the source. The photon flux from \nustar\ is likely contaminated by Kes~73 especially at $\lesssim5$~keV \citep{an13ApJ:1841, an15ApJ:1841}, yet, since we do not consider any cross-instrument spectral analysis, this contamination does not affect our results (see below and Section~\ref{sec:specres}).

The Neutron star Interior Composition Explorer \citep[\nicer;][]{gendreau16SPIE} has performed bi-weekly observations of the magnetar \src\ since 2018 as part of a long-term monitoring program for a systematic study of magnetar activity \citep[e.g.,][]{younes20ApJ:2259}. At the start of the 2024 August outburst, \nicer\ cadence was increased to closely follow-up the evolution of the source. In this paper, we report on all the 2024 \nicer\ observations which spanned the times 2024 February 13 to 2024 November 11. We create cleaned and calibrated event files using \texttt{nicerl2}, part of {\tt NICERDAS} version v12. To increase the \nicer\ data available for our analysis, we retain both night and day-time exposures with the flag {\tt threshfilter} set to ALL. Moreover, we excise any flaring intervals due to strong particle activity by visually inspecting the 12--15 keV light curve, where events are predominantly high energy particle background. We only utilize \nicer\ for timing analysis; spectral analysis with the available data is complicated by the strong contamination to the source flux from the X-ray bright Kes 73 SNR.

To increase our data statistics, we also utilize the \ixpe\ observations presented in \citet{stewart24arXiv241216036S}, primarily for timing analysis. We refer the reader to that work for details on the data reduction procedure.

We correct all photon arrival times to the solar system barycenter using the HEASoft tool \texttt{barycorr}, the JPL ephemerides DE430 and the source coordinates as established in \citet{wachter04ApJ}. Subsequently, we perform our timing analysis utilizing the Code for Rotational-analysis of Isolated Magnetars and Pulsars \texttt{CRIMP}\footnote{\url{https://github.com/georgeyounes/CRIMP/tree/main}} (version 0.1.0). Specifically, we use {\tt CRIMP} to perform period searches using the $Z^2$ statistics, perform model fitting to pulse profiles, deliver rms pulsed fraction, measure pulse times-of-arrival (TOAs), and derive the corresponding rotational phase of each photon arrival time. For TOA measurement, {\tt CRIMP} allows for a variable template model, an important feature in magnetar timing during outbursts, and relevant to the case of \src\ as we detail below. To model the pulse TOAs and create a phase coherent timing solution throughout the 2024 observational campaign, we utilize {\tt Tempo2} \citep{hobbs2006:tempo2} and {\tt PINT} \citep{luo19ascl:pint}. The two software packages resulted in near-identical timing solutions; hereafter, we report the {\tt Tempo2} results.

For spectral analysis, we utilize Xspec version 12.14.0c. We group all spectra to 5 counts per energy bin and use the W-stat (\texttt{statistic cstat} in Xspec) as our fitting statistics. We restrict the spectra to the energy range 3-70 keV since we only consider \nustar. Accordingly, we fix the absorbing column density $N_{\rm H}$ to $2.6\times10^{22}$~cm$^2$ as derived in \citet{kumar14ApJ:1841} utilizing high-resolution, soft X-ray imaging instruments. For consistency with the latter work, we adopt the Anders \& Grevesse elemental abundances (\texttt{angr} in Xspec, \citealt{anders89gca}) along with the photo-electric cross-sections of \cite{verner96ApJ:crossSect}. Lastly, we model the calibration discrepancy between the two \nustar\ detectors FPMA and FPMB with a multiplicative constant normalization, which we find to be in the $1-3\%$ range.

\subsection{Radio}
\src\ has been observed as part of two monitoring campaigns that target radio-quiet magnetars in the search for radio emission independent of X-ray outbursts. These campaigns were started in February 2024 with the MeerKAT radio telescope and in May 2024 with the Effelsberg \SI{100}{m} radio telescope. MeerKAT is a 64-dish interferometer located in the Karoo desert in South Africa, while Effelsberg is a single-dish instrument located near Bonn in Germany. 
For MeerKAT, seven epochs have been observed as part of an open time proposal with a monthly cadence while six more have been observed as part of DDT time with an increased cadence after the outburst. With Effelsberg, two observations, one each in May and June, were taken.
Beamformed search-mode data was recorded for both telescopes.
Each MeerKAT observation was taken with a time resolution of 38$\rm \mu$s, full polarization information and 1024 frequency channels. Five of the open time observation were taken at L-band (856 to 1712 MHz) and the other two and all DDT observations were taken in the S1 band (1968 to 2843 MHz).
The Effelsberg observations were taken with the new Ultra Broad Band receiver covering the frequency range from 1300 to 6000 MHz with five sub-bands (Band 1 1290 to 1940 MHz, Band 2 1940 to 2590 MHz , Band 3 2975 to 4100 MHz, Band 4 4100 to 5225 MHz, Band 5 5225 to 5975 MHz). The data was recorded using the EDD backend \citep{EDD} independently in each sub-band at a frequency and time resolution of \SI{1}{MHz} and \SI{32}{\micro s}, respectively, and full polarization information.
The duration of the observation varies by telescope and observing band. Table \ref{tab:radio_obs} gives an overview of all radio observations of \src\ presented in this work.
Given that no radio emission has been observed from \src, we estimated the amount of dispersion expected, the dispersion measure (DM), to be \SI{746}{pc.cm^{-3}} using the NE2001 model \citep{cordes2002}. This value is used for coherent de-dispersion at the time the data is taken for both telescopes, with the exception of the first Effelsberg observation, where no coherent de-dispersion has been applied.
The observations of the other sources targeted by the survey and the imaging data of MeerKAT are presented in separate, forthcoming publications.

All MeerKAT and Effelsberg data are searched for radio emission in the form of single pulses as well as a periodic (i.e., folded) profile. For the Effelsberg observations, each sub-band of the UBB is searched individually. The single pulse search is based on the software package {\tt TransientX} \citep{men2024}\footnote{\url{https://github.com/ypmen/TransientX}}, which offers inbuilt mitigation techniques for radio frequency interferences (RFI).
We searched over a DM range of \SIrange{20}{10000}{pc.cm^{-3}}, a width range from the native time resolution of the data to the period of the magnetar and applied a minimum singal to noise ration (S/N) of six. Each candidate is inspected visually and classified as real single pulse or noise/RFI.
For the search in the folded profile, we use the {\tt PRESTO} software suite \citep{ransom2001}\footnote{\url{https://github.com/scottransom/presto}}. First a mask is  generated to mitigate RFI in the data using \texttt{rfifind}. The data are then folded using the timing solution of the X-ray observations provided in this work and with 128 profile bins. The folds are thus only searching for the DM and cover the range from \SIrange{00}{22145}{pc.cm^{-3}}.

\section{Results}
\label{sec:res}

\subsection{Timing analysis}
\label{sec:timres}

\begin{figure*}[t!]
    \centering
    \includegraphics[angle=0,width=0.9\textwidth]{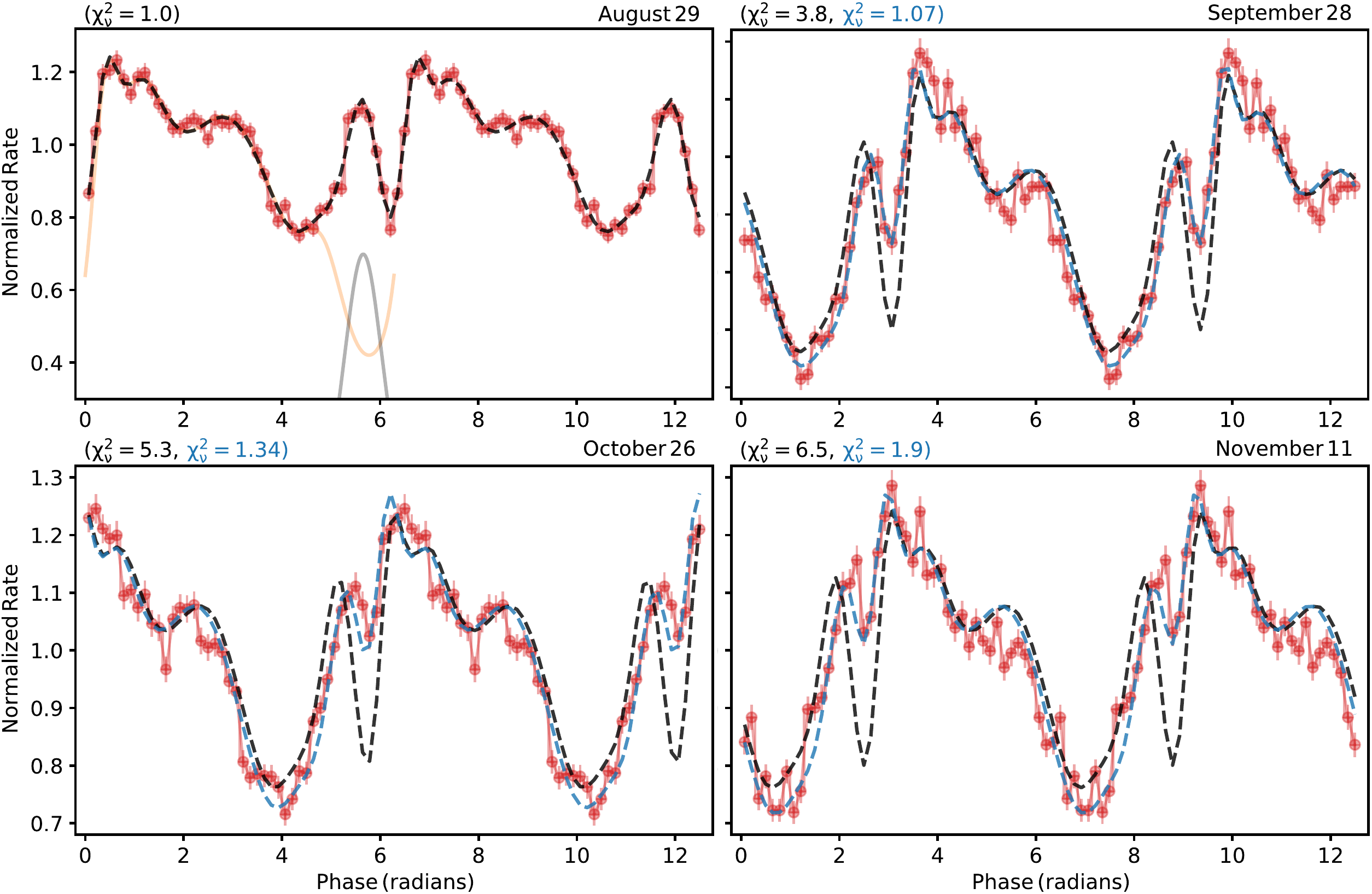}
    \caption{\nustar\ pulse profiles in the 3-10 keV range for the observations that were performed post-outburst, grouped to 45 phase-bins. Each pulse profile is derived using a local ephemerides valid only for the duration of the corresponding observation. The black-dashed line is the best-fit 5 von Mises distribution to the 2024 August 29 pulse profile (upper-left panel; gray solid line shows the von-Mises component that fits the newly-formed narrow peak, while the yellow one represents the sum of the other four), 8 days post outburst-onset. This model has been fit to the subsequent profiles, allowing for a global phase-shift and constant normalization factor. The reduced $\chi^2$ is noted in the upper-left corner of each panel. The blue-dashed line in the upper-right and bottom panels is the same as the black-line, only now the centroid of the von Mises component that fits the narrow peak is allowed to vary. The corresponding reduced $\chi^2$ are also displayed. In all panels, two cycles are plotted for clarity.}
    \label{fig:peakmigration}
\end{figure*}

\begin{figure}[t!]
    \centering
    \vspace{0.15in}
    \includegraphics[angle=0,width=0.48\textwidth]{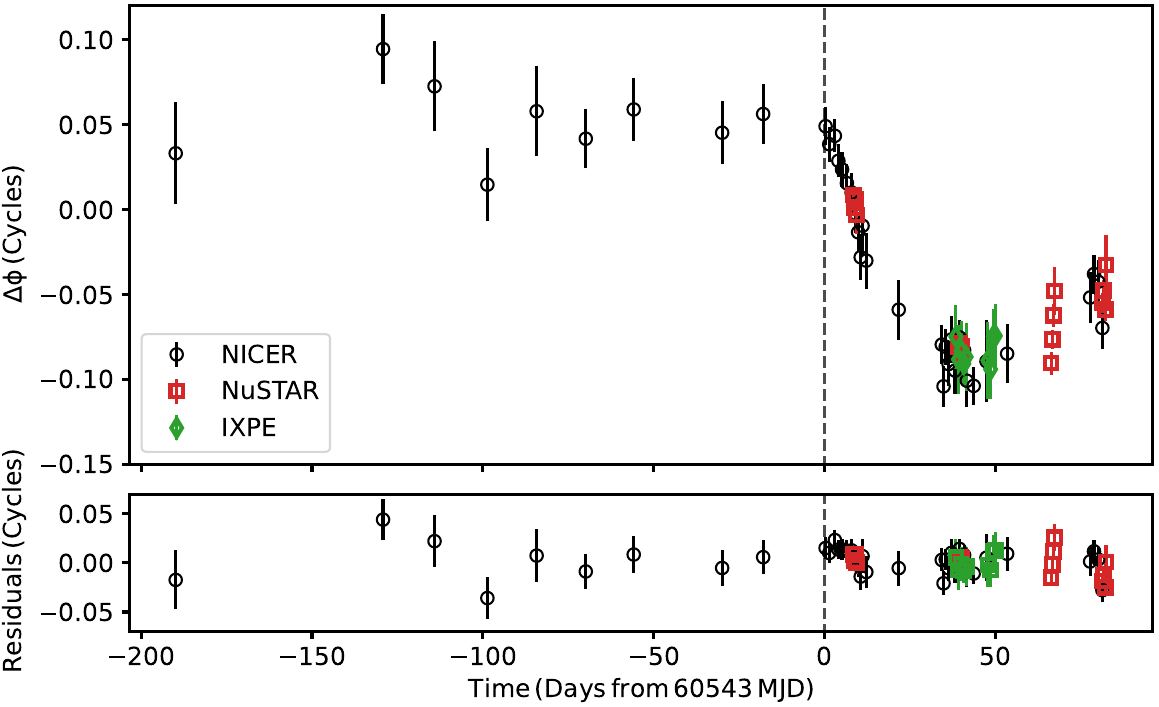}
    \caption{{\sl Upper panel.} Spin-residuals of the \src\ pulse time-of-arrival covering the full 2024 time period, after subtracting a timing model that fit the pre-outburst data (left of vertical dashed-line) consisting of a spin frequency and its first derivative. The residuals post-outburst are reminiscent of a spin-up glitch. {\sl Lower panel.} Same as above but after incorporating a glitch to the timing solution.}
    \label{fig:toaglitch}
\end{figure}

\begin{table}
\caption{Phase coherent spin parameters of \src}
\label{tab:timsol}
\vspace*{-0.25cm}
\begin{center}
\resizebox{0.45\textwidth}{!}{
\hspace*{-1.0cm}
\begin{tabular}{l r}
\hline
\hline
Right ascension (J2000) \T\B & 18:41:19.343\\ 
Declination (J2000) \T\B & $-$04:56:11.16\\ 
Time Scale \T\B & TDB \\ 
Ephemeris \T\B & DE 430 \\ 
MJD range \T\B &  60353--60625\\
Epoch (MJD) \T\B & 60450 \\
  \hline
  $\nu$ (Hz) \T\B & 0.084~702~008(1) \\
  $\dot\nu$ (Hz s$^{-1}$) \T\B &  -2.938(2)$\times10^{-13}$ \\
  $t_{\rm g}$ (MJD) \T\B & $60543$~(fixed) \\
  $\Delta\nu$ (Hz) \T\B &  $6.1(4)\times10^{-8}$ \\
  $\Delta\dot\nu$ (Hz~s$^{-1}$) \T\B & $-1.42(9)\times10^{-14}$ \\
  $\Delta\nu/\nu$ \T\B & $7.6(5)\times10^{-7}$ \\
\hline
  $\chi^2$/dof \T\B &  89/58\\
   RMS residual (ms) \T\B &  188 \\
  \hline
\hline
\end{tabular}}
\end{center}
{\bf Notes.} The 1 $\sigma$ uncertainty on each parameter is given in parentheses.
\end{table}

\begin{figure*}[t!]
\begin{center}
  \includegraphics[angle=0,width=0.99\textwidth]{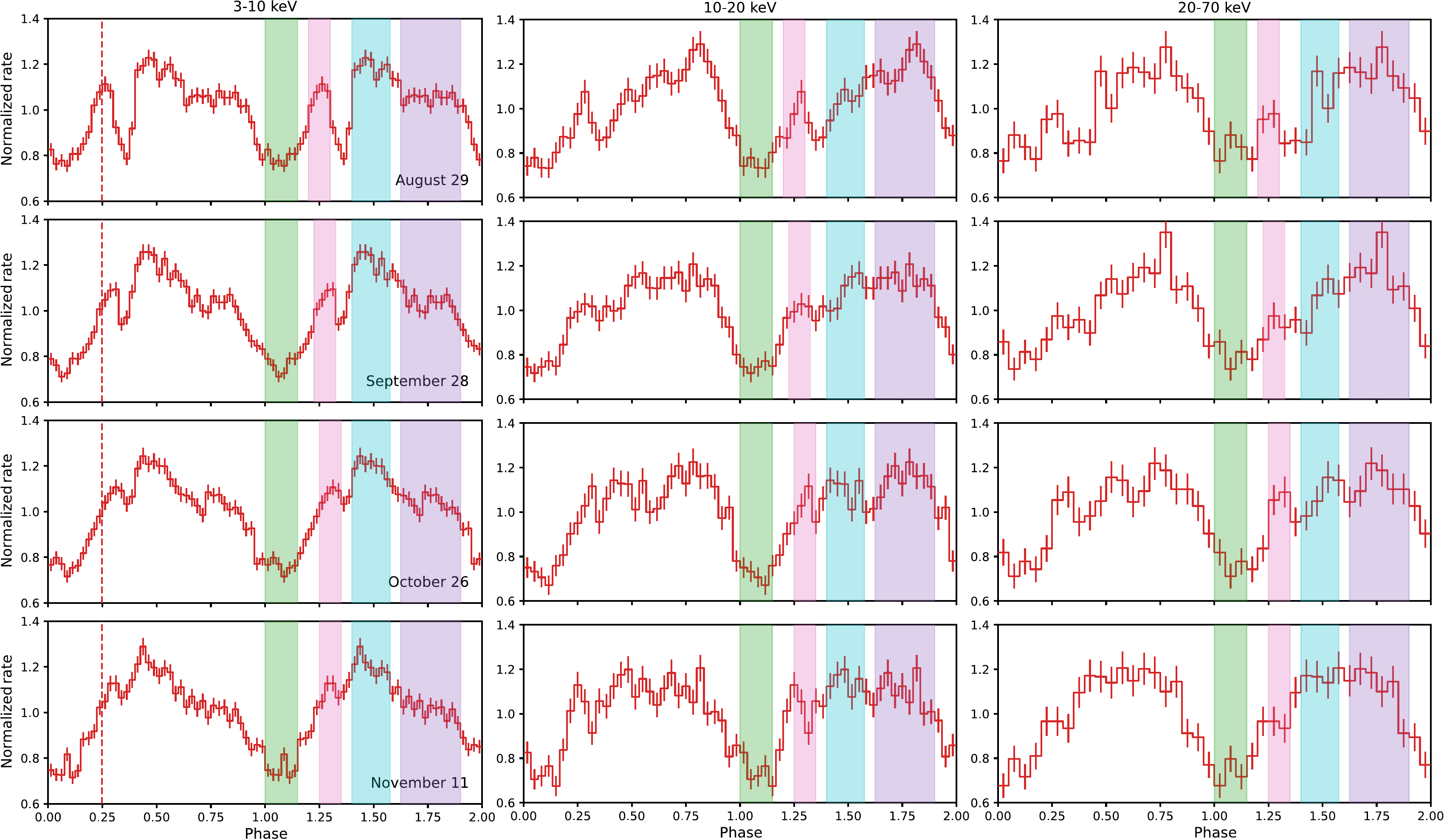}
\caption{Energy-resolved pulse profiles as observed with \nustar\ post-outburst. Each row represents a single observation, while columns show the pulse-profiles in increasing energy bins. Data folded with the full phase-coherent timing solution covering the entirety of the 2024 datasets (Table~\ref{tab:timsol}). The shaded areas delimitate the intervals we used for phase-resolved spectroscopy. The dashed line in the 3-10 keV pulse-profiles represents the centroid of the best-fit von Mises component to the newly-formed peak as derived for the earliest \nustar\ observation.}
\label{fig:nusprof}
\end{center}
\end{figure*}

Our preliminary timing analysis upon the acquisition of the earliest post-outburst \nicer\ and \nustar\ data (taken between August 21 and August 29) reveals a prominent change to the soft X-ray pulse profile shape, namely the appearance of a narrow peak with a width of about 0.1~cycles. The peak was also present in the later observations, yet it showed a noticeable shift towards merger with the main pulse. To assess the significance of this shift, we utilize the \nustar\ observations due to their compactness and nice spread over the following three months, prior to the source entering Sun avoidance angles (Table~\ref{tab:specres}). We measure the local spin-frequency of each \nustar\ observation utilizing the $Z^2$ method with number of harmonics $n=5$ in the energy range 3-50 keV. While these choices maximized the $Z^2$ power, our results are not affected by them. We observe a strong signal with a peak $Z^2\gtrsim1000$ in each of the observation corresponding to the spin-frequency of the magnetar. We refine this measurement through a phase-coherent analysis (as detailed below) and we derive a consistent uncertainty on the spin-frequency of about $1\times10^{-7}$~Hz across all observations. 

We fold the arrival time of the soft X-ray events (3-10 keV) of each observation utilizing the above local ephemerides and group the resulting phases to 45 bins. These pulse-profiles are shown in Figure~\ref{fig:peakmigration}. We fit the profile of the earliest \nustar\ observation to an increasing number of von-Mises (wrapped Gaussian) functions and assess the quality of the fit according to the $\chi^2$ test. We find that a model consisting of 3 von-Mises components or less cannot satisfactorily fit the data resulting in a reduced $\chi^2\gg2$. Four von-Mises components provide an acceptable fit to the data with a reduced $\chi^2$ of 1.6. Adding one extra component results in a reduced $\chi^2$ of approximately 1, and an improvement at the $3.5\sigma$ level according to an F-test. This fit is shown as a black dashed-line in the upper-left panel of Figure~\ref{fig:peakmigration}. We then fit this model to the later \nustar\ observations allowing only a global phase-shift and a global normalization constant. These are also shown as black dashed-lines in the other three panels, along with their corresponding reduced $\chi^2$. We then allow 1 extra parameter to freely vary, the centroid of the von-Mises function that describes the newly formed peak (gray solid-line in upper-left panel of Figure~\ref{fig:peakmigration}). The resultant best-fit models are shown as blue dashed-lines, and their corresponding $\chi^2$ also displayed. The improvement in the fit is highly statistically significant in all three cases, while the new best-fit centroid imply a  rightward shift with time. As an extra layer of checks, we repeat the analysis with the model consisting of 4 von-Mises distribution and find consistent results. We conclude that, indeed, the newly formed peak is exhibiting a shift in rotational phase as a function of time, towards a merger with the main pulse. We incorporate this shift in the remainder of our analyses whenever necessary.

\begin{figure}[t!]
    \centering
    \hspace{-0.2in}
    \includegraphics[width=0.47\textwidth]{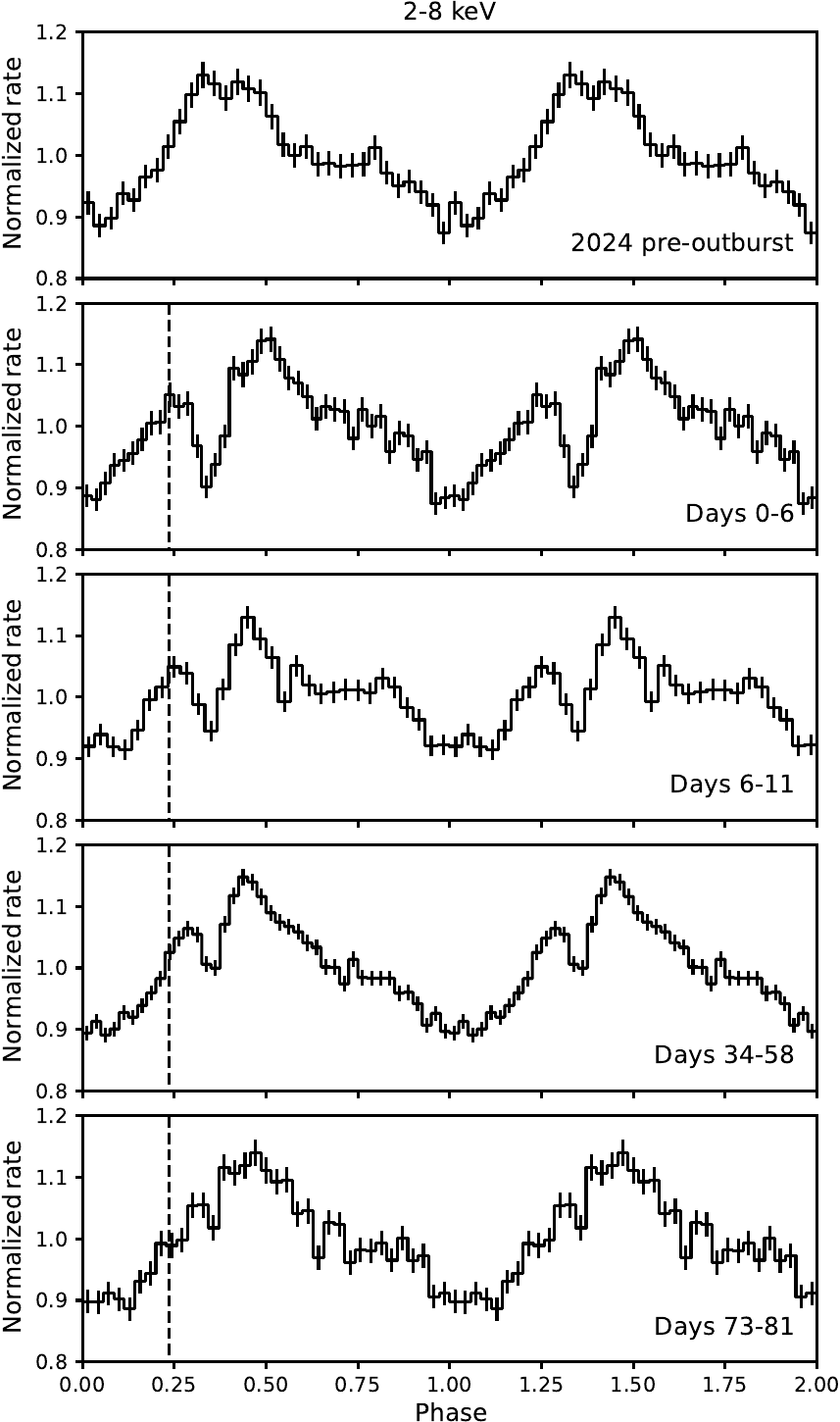}
    \caption{\nicer\ pulse profiles of \src\ as observed pre-outburst (upper-panel) and post outburst. Day 0 corresponds to 2024 August 21. Data folded with the full phase-coherent timing solution covering the entirety of the 2024 datasets. The dashed line in the bottom four pulse-profiles represents the centroid of the best-fit von Mises component to the newly-formed peak as derived for the earliest \nicer\ data (days 0-6).}
    \label{fig:nicerprof}
\end{figure}

We rely on the long-term \nicer\ monitoring of \src\ to perform our phase-coherent timing study. For the purposes of this paper, we only report on the 2024 data, for which we establish a phase-coherent timing solution from February to the August 2024 outburst onset. We first build a template by fitting a pre-outburst high S/N \nicer\ profile to a Fourier series with 3 harmonics. We then establish the time-intervals that define each TOA by merging consecutive $2\times10^4$ \nicer\ events detected within $<2$-day period. Finally, we fit the unbinned, folded profile of each of these time-intervals to the same template only allowing a global phase-shift $\Delta\phi$ and a normalization constant to vary, utilizing the extended maximum likelihood fitting method with a Poisson probability density function. We find that a simple timing model consisting of a spin frequency and its first derivative is adequate to fit the phase-shift, or equivalently, the pulse TOA with a $\chi^2$ of 9 for 7 dof. The resulting root-mean-squared (rms) residuals is 130~ms ($1.3\%$ of the source spin cycle).

Given the drastic change in the source pulse shape post-outburst, we utilize the best-fit von-Mises model to the first \nustar\ observation as template. Similar to above, we construct time-intervals that define each post-outburst ToA by merging consecutive $2\times10^4$ \nicer\ events ($10^4$ for \nustar\ and 5000 for \ixpe) detected within $<2$-day period. We fold these events with the timing solution established above, and fit the resulting profiles to the template while allowing a phase-shift $\Delta\phi$, a normalization constant, and the centroid of the newly-formed peak to vary (we restrict the latter to only vary between -0.1 and 0.1 cycles). The residuals of these ToAs are presented in Figure~\ref{fig:toaglitch}. The sudden, and quasi-linear shift in residuals, followed by a quadratic turn is consistent with a spin-up glitch. Indeed, fitting the full 2024 ToAs to a model including a glitch in frequency and its derivative provides a good fit to the data with a reduced $\chi^2$ of 1.5 for 58 dof, and an rms residuals of 188~ms\footnote{The timing model also includes a global phase-shift that is uniformly applied only to the post-outburst TOAs which minimizes any abrupt jump in the residuals due to (1) the different template used between pre- and post-outburst TOAs, and (2) the fixed glitch epoch. This global shift is $0.015\pm0.011$~cycles.}. We summarize our timing fit in Table~\ref{tab:timsol}. We find $\Delta\nu =(6.1\pm0.4)\times10^{-8}$~Hz and $\Delta\dot{\nu}=(-1.4\pm0.1)\times10^{-14}$~Hz~s$^{-1}$. We note that we fixed the glitch epoch to the first reported BAT burst in August, i.e., MJD 60543.

We display the folded \nustar\ pulse profiles with the full phase-coherent solution, in three separate energy ranges, in Figure~\ref{fig:nusprof}. In addition to the noticeable change at soft X-rays, the pulse also exhibits variability at hard X-rays. The usually flat-topped pulse profile at energies $>10$~keV \citep[e.g.,][]{an13ApJ:1841} shows a more prominent peaked-profile with maximum at around phase 0.75. The high energy pulse returns to its pre-outburst shape in the last \nustar\ observation of November 11. We also show the folded 2-8 keV \nicer\ data in Figure~\ref{fig:nicerprof} which present a consistent picture compared to the \nustar\ data in the 3-10 keV range. Finally, we measure the rms pulse fraction (PF) of each of these profiles. We do not find strong variation with time; the \nustar\ rms PF is 11\%, 13\% and 15\%\ in the 3-10, 10-20, and 20-70 keV, respectively, while the \nicer\ PF is around 6\% throughout the outburst (smaller than \nustar\ partly due to the strong contamination from the bright SNR Kes 73). These values are also consistent with the pre-outburst measurements.

\begin{figure}[t!]
    \centering
    \hspace{-0.2in}
    \includegraphics[width=0.49\textwidth]{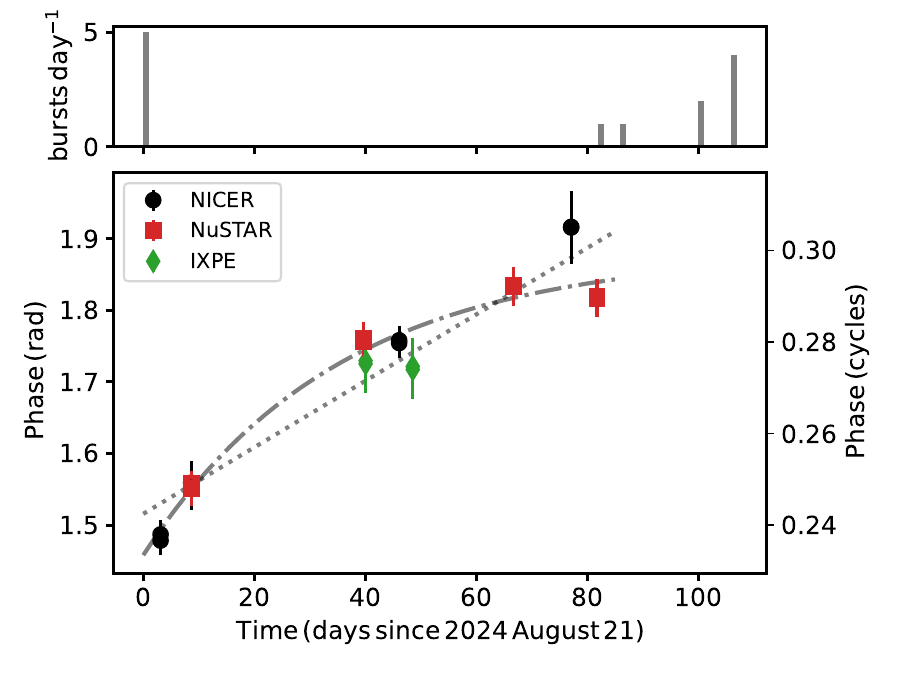}
    \caption{{\sl Upper panel.} Daily burst history of \src\ in 2024 and up to February 2025 as observed with Swift-BAT. {\sl Lower panel.} Pulse phase, in radians, of the newly formed peak in the soft X-ray band as a function of time. The dotted line is a linear fit to the data, while the dotted-dashed line is an exponential decay in the increasing form.}
    \label{fig:pulsemig}
\end{figure}

Finally, to quantify the phase-shift of the newly-formed peak, we replicate the analysis which we detailed at the beginning of this section, except now we utilize the full phase-coherent solution. In summary, we fit each \nicer, \nustar, and \ixpe\ profile (presented in Figures~\ref{fig:nusprof}, \ref{fig:nicerprof}, and in \citealt{stewart24arXiv241216036S}) to a von-Mises model that best fits the first \nustar\ profile. We allow the centroid of the newly formed peak (marked as dashed lines in Figure~\ref{fig:nusprof} and \ref{fig:nicerprof}), the overall normalization, and the overall pulsed-fraction (i.e., one multiplicative constant normalization applied to the amplitude of each of the von-Mises components) to vary. The latter takes into account the differing level of background in the different detectors which impacts the level of pulsed flux observed. Finally, we repeated the analysis utilizing 4 and 5 von-Mises components which returned consistent results. We show the phase-shifts from the latter fits in Figure~\ref{fig:pulsemig}. There is a clear trend of an increasing pulse phase with time. We fit this evolution to a linear function which resulted in a slope of $4.6(3)\times10^{-3}$~rad~day$^{-1}$, or $7.3(5)\times10^{-4}$~cycle~day$^{-1}$, and a reduced $\chi^2$ of 3.1 for 8 dof (dotted line). We also fit the evolution to an exponential-decay function of the increasing form $a(1-\exp(-t/\tau))+norm$, where $t$ is the time in days, $\tau$ is the exponential decay timescale also in days, and $norm$ the initial phase of the newly formed peak at $t=0$ (MJD 60543). We find $a=0.45_{-0.06}^{+0.08}$~rad, $\tau=43_{-13}^{+20}$~days, and $norm=1.45\pm0.02$~rad (dot-dashed line). The fit results in a reduced $\chi^2$ of 1.0 for 7 dof, which is a slight improvement compared to the linear fit. We also present in Figure~\ref{fig:pulsemig} the 2024 to February 2025 burst history of \src\ as observed with Swift-BAT \citep{palmer24ATel16927}.

\subsection{Spectral analysis}
\label{sec:specres}

We solely rely on the \nustar\ observations to present the spectral evolution of \src\ during its 2024 August outburst, mainly due to its nearly equal spread over the first 3 months of outburst and its hard spectral coverage ($>3$~keV) which limits the SNR contribution to the point source flux. As a baseline, we analyze the historical 2012 \nustar\ observation (id 30001025002), which has a similar exposure of about 50~ks to the post-outburst datasets. We fit the spectrum of this latter observation to an absorbed double power-law (2PL) model with a fixed $N_{\rm H}$  (see Section~\ref{sec:obs}). This model provides a statistically good fit with a W-stat of 1400 for 1464 dof. Adding another component, e.g., a blackbody (BB) does not significantly improve the quality of the fit with a W-stat of 1390 for 1462 dof. We also test the validity of a blackbody (BB) and a PL fit to the data, yet this combination results in a worse fit with a W-stat of 1507 for the same dof. Lastly, neither a broken-PL nor an exponential-cutoff PL provide a satisfactory fit to the data. Hence, we deduce that a 2PL model is an adequate representation of the \nustar\ spectra pre-outburst. We note that prior analyses of the broadband X-ray spectrum of \src\ indicate that the emission at energies $>3$~keV is dominated by non-thermal components \citep{an13ApJ:1841, an15ApJ:1841, enoto17ApJS}. 

\begin{figure*}[t!]
    \centering
    \hspace{-0.2in}
    \includegraphics[width=0.47\textwidth]{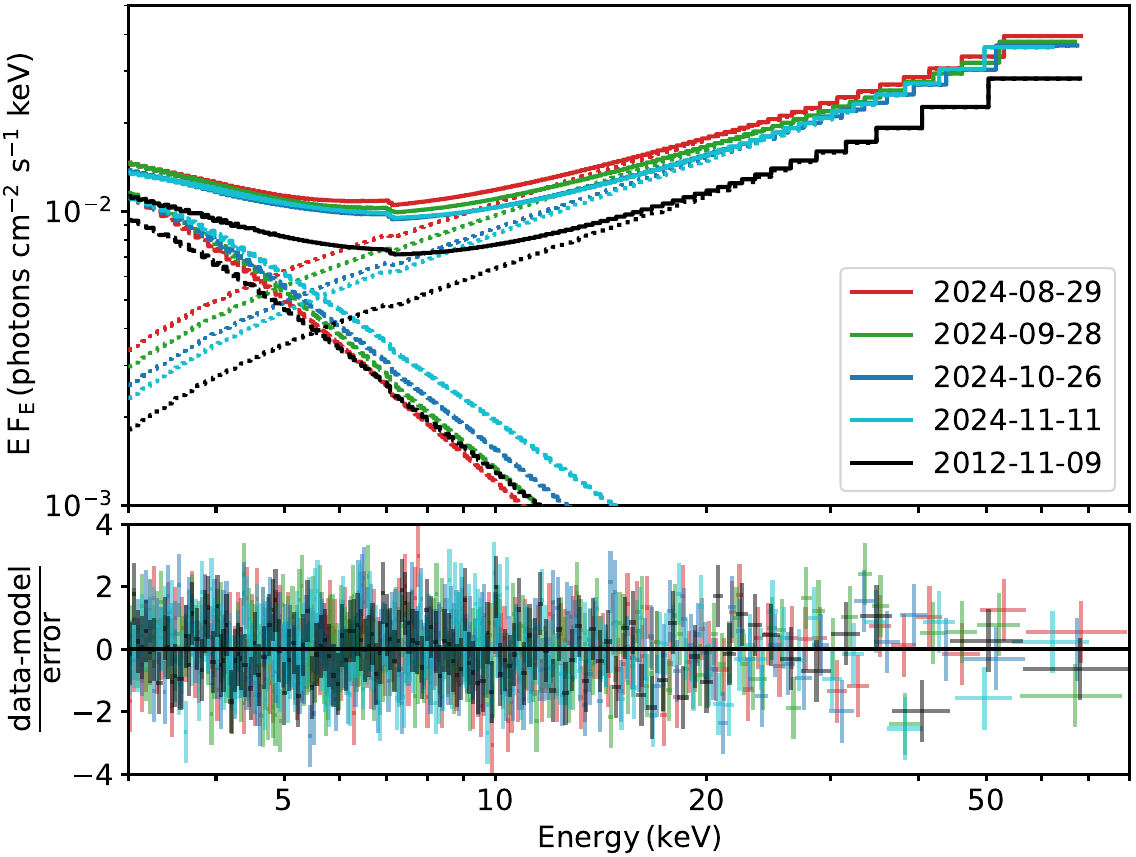}
    \includegraphics[width=0.47\textwidth]{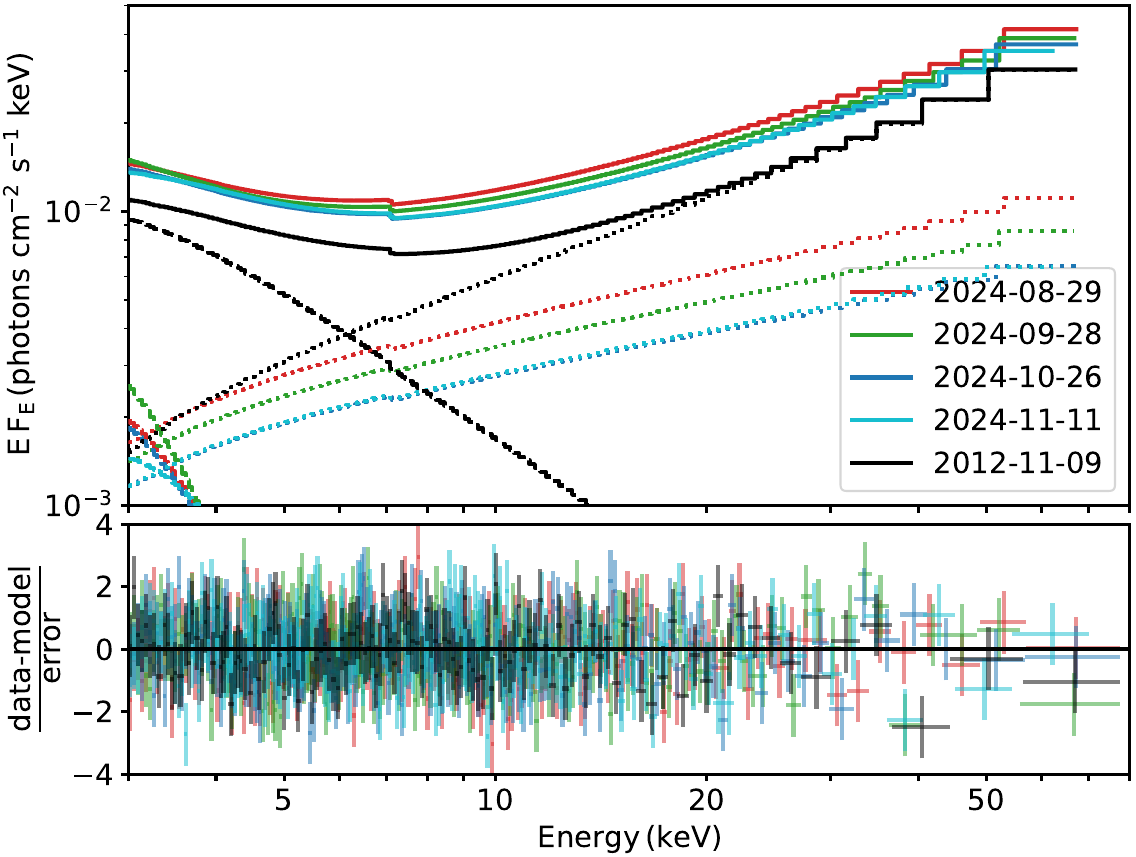}
    \caption{{\sl Upper-left panel.} Best-fit 2PL model to pre-outburst \nustar\ data (black) and post-outburst \nustar\ spectra (colored lines), shown in $\nu F_{\rm \nu}$ space. The total model flux is shown as a solid line while individual components are shown as dashed (SPL) and dotted (HPL). {\sl Lower-left panel}. Residuals of the fit in units of 1 standard deviation. {\sl Upper-right panel.} Best-fit 2PL model to pre-outburst \nustar\ data (black) while adding a BB+PL model to the post-outburst \nustar\ spectra (colored lines), shown in $\nu F_{\rm \nu}$ space. The total model flux is shown as a solid line while individual components are shown as dashed (SPL for the pre-outburst and BB for the post-outburst) and dotted (HPL). {\sl Lower-right panel}. Residuals of the fit in units of 1 standard deviation.}
    \label{fig:spec}
\end{figure*}

We then fit the 2PL model to each of the post-outburst spectra, which we also find to be an adequate spectral decomposition to the data. We present these results in Table~\ref{tab:specres}, while the best-fit model and residuals in terms of $\sigma$ are shown in Figure~\ref{fig:spec}. We find largely similar photon indices when comparing post- and pre-outburst spectral fits. On the other hand, the source flux 1 week from outburst onset is about 60\% larger than the baseline. This increase is predominantly observed in the hard X-rays, where the hard X-ray tail contributes. The flux declines by about $10\%$ over the course of our \nustar\ monitoring.

We also investigate adding the pre-outburst best-fit model to the spectral analysis of the post-outburst spectra, assuming that the former is unchanged during outbursts. In this scenario, fitting the post-outburst enhancement to a 2PL results in a soft component with a photon index $\Gamma\approx5$, which implies a thermal-nature for the soft X-ray enhancement. Hence, we also fit the data with a BB+PL model (Table~\ref{tab:specres}). We find a blackbody temperature of about 0.43 keV and an average effective emitting area of about $36$~km$^2$ adopting an 8.5~kpc distance. We find an average PL photon index $\Gamma\approx1.5$, which is somewhat softer than the persistent hard X-ray tail. We observe a significant flux decrease in this non-thermal component between the first and last \nustar\ observation from $2.7\times10^{-11}$~erg~s$^{-1}$~cm$^{-2}$ to $1.9\times10^{-11}$~erg~s$^{-1}$~cm$^{-2}$. We do not identify any strong variability in the spectral curvature, however.

Lastly, we perform phase-resolved spectroscopy to identify the spectral properties of each of the features in the source pulse profile. Our pulse phase binning is presented in Figure~\ref{fig:nusprof}, where the green shaded area represents the off-pulse spectrum (considered as background), while the pink, blue, and purple shaded areas correspond to the newly-formed peak, the main soft X-ray pulse, and its trailing edge, respectively, and are referred to as intervals 1, 2, and 3 in Table~\ref{tab:pulsedspecParam} (note that the pink shaded area is slightly shifted rightward in the later \nustar\ data to accommodate the pulse-peak shift we identify in Section~\ref{sec:timres}). We find that the pulsed spectra in each of these intervals are best fit with a non-thermal PL model; a BB model results in temperatures of about 2~keV, too high to correspond to surface thermal emission. Moreover, we find that a second component is not required by the data. We summarize the photon indices for each peak during each of the \nustar\ post-outburst observations in Table~\ref{tab:pulsedspecParam}. We find a consistent PL index between intervals 1 and 2, while interval 3 exhibits a harder PL. Moreover, the latter softens with time. This is consistent with the picture emerging from our energy-resolved pulse profile analysis, which indicates a brighter emission at hard X-rays in the trailing edge of the main pulse at the early times of the outburst.

\begin{table*}[t]
\caption{Best fit spectral parameters of \nustar\ observations}
\label{tab:specres}
\vspace*{-0.6cm}
\begin{center}{
\resizebox{\textwidth}{!}{
\hspace*{-2.5cm}
\begin{tabular}{c c c c c c c c c c}
\hline
\hline
Observation ID   \T\B & Date & Exposure & $kT$ & $R_{\rm BB}^{2*}$ & $\Gamma_{\rm S}$ & $\Gamma_{\rm H}$ & $F_{\rm S}^{**}$ & $F_{\rm H}^{**}$ & $F_{\rm tot}^{**}$ \\
                 \T\B & (UT) & (ks)   & (keV) & (km$^2$) & & & & & \\
\hline
\multicolumn{10}{c}{All observations treated separately \T\B}\\
\hline
30001025002 \T\B & 2012-11-09 & 49 & \ldots & \ldots & $3.9\pm0.1$ & $1.32\pm0.04$ & $2.53\pm0.06$ & $5.7\pm0.1$ & $8.24\pm0.08$\\
91001330002 \T\B & 2024-08-29 & 50 & \ldots & \ldots & $4.1\pm0.1$ & $1.32\pm0.03$ & $3.00\pm0.05$ & $8.8\pm0.1$ & $11.7\pm0.1$\\
91001335002 \T\B & 2024-09-28 & 56 & \ldots & \ldots & $4.1\pm0.1$ & $1.29\pm0.03$ & $3.07\pm0.06$ & $8.2\pm0.1$ & $11.3\pm0.1$\\
81002306002 \T\B & 2024-10-26 & 53 & \ldots & \ldots & $3.9\pm0.1$ & $1.26\pm0.03$ & $3.0\pm0.1$ & $7.6\pm0.1$ & $10.6\pm0.1$\\
81002306004 \T\B & 2024-11-10 & 51 & \ldots & \ldots & $3.7\pm0.1$ & $1.21\pm0.04$ & $3.07\pm0.06$ & $7.6\pm0.1$ & $10.7\pm0.1$\\
\hline
\multicolumn{10}{c}{Post-outburst spectral parameters when including the pre-outburst quiescent contribution \T\B}\\
\hline
91001330002 \T\B & 2024-08-29 & 50 & $0.43\pm0.05$ & $28_{-17}^{+48}$ & \ldots & $1.50\pm0.04$ & $0.4\pm0.1$ & $2.7\pm0.2$ & $3.1\pm0.2$ \\
91001335002 \T\B & 2024-09-28 & 56 & $0.38\pm0.05$ & $85_{-55}^{+200}$ & \ldots & $1.53\pm0.05$ & $0.6\pm0.2$ & $2.4\pm0.3$ & $3.0\pm0.3$\\
81002306002 \T\B & 2024-10-26 & 53 & $0.43\pm0.05$ & $26_{-16}^{+52}$ & \ldots & $1.56\pm0.07$ & $0.4\pm0.1$ & $1.8_{-0.2}^{+0.4}$& $2.2\pm0.3$\\
81002306004 \T\B & 2024-11-10 & 51 & $0.53\pm0.08$ & $5_{-3}^{+10}$ & \ldots & $1.55\pm0.08$ & $0.27\pm0.05$ & $1.9\pm0.3$& $2.2\pm0.3$\\
\hline
\hline
\end{tabular}}}
\end{center}
\begin{list}{}{}
\item[{\bf Notes.}]$^{*}$Derived by adopting an 8.5~kpc distance. $^{**}$ Derived in the 2-70~keV range in units of $10^{-11}$~erg~s$^{-1}$~cm$^{-2}$. Listed uncertainties are at the $1\sigma$ level.
\end{list}
\end{table*}

\begin{table}
\caption{Phase-resolved pulsed spectral modeling}
\label{tab:pulsedspecParam}
\vspace*{-0.6cm}
\begin{center}
\resizebox{0.5\textwidth}{!}{
\hspace*{-1.6cm}
\begin{tabular}{l c c c}
\hline
\hline
Date \T\B & Interval 1 & Interval 2 & Interval 3 \\
    (UT)           \T\B &  \multicolumn{3}{c}{$\Gamma$}\\
\hline
2024-08-29 \T\B & $2.2\pm0.2$ & $2.2\pm0.1$ & $1.33\pm0.05$\\
2024-09-28 \T\B & $1.7\pm0.2$ & $2.3\pm0.1$ & $1.39\pm0.05$\\
2024-10-26 \T\B & $1.8\pm0.1$ & $2.1\pm0.1$& $1.42\pm0.06$\\
2024-11-10 \T\B & $2.1\pm0.2$ & $2.1\pm0.1$& $1.56\pm0.06$\\
\hline
\hline
\end{tabular}}
\end{center}
{\bf Notes.} The intervals refer to the shaded areas in figure~\ref{fig:nusprof}; interval 1 for pink, interval 2 for cyan, and interval 3 for purple. Observation ids and exposures are reported in Table~\ref{tab:specres}.
\end{table}

\subsection{Radio upper limits}
After reducing the data from all observations, no single pulse or pulsed signal was found. Hence, our observations provide a time series of upper limits on the radio flux density before and after the outburst shown in Table~\ref{tab:radio_obs}.
To estimate the upper limits for the flux density of folded profile $S_{\rm{mean}}$, as well as the fluence of single pulses $F_{\rm{SP}}$, we use the radiometer equation as derived in \cite{lorimer2004} and derived the limit for $F_{\rm{SP}}$:
\begin{align*}
S_{\rm{mean}} &= \frac{(S/N) T_{\rm{sys}}}{G \sqrt{n_{\rm{pol}} t_{\rm{obs}} B}} \sqrt{\frac{X}{1-X}} \\
F_{SP} &= \frac{(S/N) T_{\rm{sys}} \sqrt{w}}{G \sqrt{n_{\rm{pol}} B}},
\end{align*}

where $(S/N)$ is the signal to noise ratio of the folded profile or single pulse, $T_{\rm{sys}}$ is the system temperature, $G$ is the gain of the receiving system, $n_{\rm{pol}}$ is the number of polarization summed (always 2 for our observations), $t_{\rm{obs}}$ is the duration of the observation, $B$ is the bandwidth of the receiving system, $X$ is the duty cycle of the magnetar and $w$ is the pulse width of the single pulse.

For estimating the upper limits we assume fiducial values of $X$ = 10\% (folded profile) and $w$ = 1~ms (single pulse), but other duty cycles and widths can easily be adopted by scaling the limits accordingly. The minimal S/N of 6 for a single pulse follows directly from the single pulse search. To find the minimal $S/N$ of the folded profile that we could have detected, we investigate the $S/N$ levels of residual RFI contributions. For MeerKAT, the RFI contribution is very low after the mitigation applied, making detections down to a $S/N$ of 5 possible in each observation, which would make the signal significantly brighter than the RFI pulses in the data (except the observation on 23 September, which is heavily affected by RFI and thus not considered). For the L-band observations, we adopt the sensitivity from \cite{chen2023} ($G$ = \SI{2.65}{K/Jy} and $T_{\rm{sys}}$ = \SI{26}{K}), while for S-band we estimate the SEFD = $T_{\rm{sys}}/G$ for 56 antennas to be \SI{8.6}{Jy}\footnote{\url{https://skaafrica.atlassian.net/wiki/spaces/ESDKB/pages/277315585/MeerKAT+specifications}}. For Effelsberg, the RFI situation is worse. Here, we estimate the minimal detectable $S/N$ of a folded profile to be 8. For Effelsberg, $G$ is \SI{1.3}{K.Jy^{-1}} for each sub-band, while $T_{sys}$ of each band are \SI{20.6}{K}, \SI{16.1}{K}, \SI{16.5}{K}, \SI{18.8}{K} and \SI{19}{K}\footnote{\url{https://eff100mwiki.mpifr-bonn.mpg.de/doku.php?id=information_for_astronomers:rx:p170mm}} respectively. We note that our upper limit estimates to pulsed emission do not take into account the contribution of red noise, which can increase the minimum detectable flux density by a factor of a few \citep[e.g.][]{2015ApJ...812...81L}. Quantifying this requires injecting synthetic pulses into the data, which is beyond the scope of this paper.

\begin{table*}
    \centering
        \caption{Overview of the radio observations of \src. For each observation, the date, MJD of the topocentric start of the observation, duration of the observation ($t_{obs}$), telescope name and observing band, the limit on the mean flux density $S_{mean}$ and single pulse fluence $F_{SP}$ are given.}
    \label{tab:radio_obs}
    \begin{tabular}{rccccc}
    \hline
    \hline
        Obs. date & MJD & $t_{obs}$/s & Telescope & $S_{mean}$/$\rm \mu$Jy & $F_{SP}$/mJy~ms\\
        \hline
         2024-02-03 & 60343.345 & 730 & MKT L-band & 17 & 52\\
         2024-03-03 & 60372.265 & 668 & MKT S-band & 13 & 39\\
         2024-04-05 & 60405.159 & 730 & MKT L-band & 17 & 52\\
         2024-05-01 & 60431.033 & 730 & MKT L-band & 17 & 52\\
         2024-05-09 & 60439.143 & 1200& EFF UBB B1 & 34 & 84\\
                    &           &     & EFF UBB B2 & 27 & 66\\
                    &           &     & EFF UBB B3 & RFI & 51\\
                    &           &     & EFF UBB B4 & 24 & 58\\
                    &           &     & EFF UBB B5 & 30 & 72\\
         2024-06-01 & 60462.023 & 730 & MKT L-band & 17 & 52\\
         2024-06-12 & 60473.144 & 1200& EFF UBB B1 & 34 & 84\\
                    &           &     & EFF UBB B2 & 27 & 66\\
                    &           &     & EFF UBB B3 & 21 & 51\\
                    &           &     & EFF UBB B4 & 24 & 58\\
                    &           &     & EFF UBB B5 & 30 & 72\\
         2024-07-12 & 60503.881 & 924 & MKT S-band & 11 & 39\\
         2024-08-28 & 60550.724 & 922 & MKT L-band & 15 & 52\\
         2024-09-17 & 60570.796 & 1200& MKT S-band & 10 & 39\\
         2024-09-22 & 60575.681 & 1200& MKT S-band & 10 & 39\\
         2024-09-23 & 60576.871 & 1200& MKT S-band & RFI &39\\
         2024-09-28 & 60581.765 & 1200& MKT S-band & 10 & 39\\
         2024-10-11 & 60594.712 & 1200& MKT S-band & 10 & 39\\
         2024-10-28 & 60611.651 & 1200& MKT S-band & 10 & 39\\
         \hline
         \hline
    \end{tabular}

\end{table*}

\section{Discussion}
\label{sec:disc}

\subsection{High energy outburst}

We have presented the timing and spectral evolution of the magnetar \src\ during the first 82~days of its 2024 outburst \citep{dichiara24ATel16784}. We observe a clear increase in flux, by about 20\% at soft energies and 60\% at hard X-rays, with minimal decay. The flux enhancement is predominantly driven by the presence of a hard X-ray component that is well fit with a PL having a photon index $\Gamma\approx1.5$. In magnetar outburst theory, this component is typically attributed to newly-formed, magnetospheric twisted B-field loops driven by plastic motion of the corresponding crustal foot-points \citep{thompson02ApJ:magnetars}. These twisted field bundles are capable of accelerating particles to high Lorentz factors. Bulk surface thermal photons with temperatures of about 0.5 keV can resonantly up-scatter off of these accelerated particles, resulting in inverted hard X-ray spectra \citep{baring07, fernandez07ApJ, wadiasingh18ApJ}. Additionally, if the particles' Lorentz factor exceeds $\sim 10$, single photon pair-creation could ignite, resulting in pair cascades \citep[][Harding et al. in prep.]{baring01ApJ,2019MNRAS.486.3327H,2019BAAS...51c.292W}. These pairs, if created in high Landau states at higher altitudes, emit synchrotron emission from the outer parts of equatorial field loops. The highly polarized hard X-ray tail recently reported in \cite{stewart24arXiv241216036S,rigoselli24arXiv241215811R} indeed suggests that synchrotron emission might play a significant role in shaping the spectrum of the hard X-ray emission \citep{stewart24arXiv241216036S}. Surface bombardment by return currents from twisted field loops could accommodate the increase that is observed at soft X-rays. Finally, the modest enhancement that is exhibited by \src\ is largely due to its bright X-ray persistent state \citep{an15ApJ:1841}, and consistent with the maximal soft and hard X-ray luminosity expected from magnetar surfaces and magnetosphere, respectively. For the surface, the electromagnetic radiation is regulated through neutrino emission and cannot exceed $\sim 10^{36}$~erg~s$^{-1}$ whereas hard X-rays from a maximally-twisted, axisymmetric field loop could reach a similar luminosity \citep{pons12ApJ:mag,beloborodov09ApJ}.

The spin-up glitch observed in \src\ at outburst onset is an almost ubiquitous feature of the continuously-monitored members of the X-ray magnetar family \citep[e.g.,][]{dib14ApJ}. Its size and the sudden change to $\Delta\dot{\nu}$ are also on-par with other measurements \citep[e.g.,][]{dib14ApJ,hu24Natur}. In standard pulsar glitch theory, spin-up glitches occur when the superfluid component of the inner crust transfers part of its larger angular momentum to the outer-crust, which suffers from magnetic-dipole braking \citep[see][and references therein]{haskell15IJMPD,2022RPPh...85l6901A}. Under this assumption, the fact that the majority of magnetar outbursts from the monitored bunch are accompanied by glitching activity may indicate that their trigger mechanism is intimately connected to the inner crust. Yet, the accompanying and profound radiative changes, which is rarely observed in canonical rotation-powered pulsars \citep{palfreyman18Natur}, argue for a link between internal instabilities causing the glitch and crustal and magnetospheric energy deposition that lead to the observed outbursts. It is worth noting that spin-up glitches can also occur without any obvious signs of activity such as the emission of short bursts \citep[e.g.,][]{dib14ApJ, younes20ApJ:2259}. It is unclear whether this attribute is intrinsic to the population or due to some observational bias such as our incompleteness to sample the full magnetar burst luminosity function, especially with the current large field-of-view missions such as Swift-BAT and Fermi-GBM \citep{collazzi15ApJS, younes20ApJ1935, lin20ApJ:1935}.

Arguably, the most intriguing result of our \src\ investigation is the apparent shift of the source newly-formed pulse-peak, the second observation of its kind, following SGR~1830$-$0645 \citep{younes22ApJ:ppm}. The maximum possible rate is that occurring along the great circle at the equator of a sphere with a $\sim10$~km radius (observed edge-on with respect to the spin axis), the speed translates to $\lesssim50$~m~day$^{-1}$, about half the rate of SGR~1830$-$0645. Yet, this relatively small difference can easily be accounted for when considering the large systematic uncertainties due to the exact shape of the emitting regions, the full geometry of the system, and the compactness of the stars. The two observations share other similarities as well. For instance, similar to SGR~1830$-$0645, the shift in \src\ is linear in time, and in the direction to simplify the pulse shape (a common trait of magnetars \citealt[e.g.,][]{gavriil11ApJ:0142,scholz12ApJ:1822}). Moreover, the newly-formed peak is narrow with a width of about 0.1~cycles, comparable to the width of the shifting peaks in SGR~1830$-$0645. Yet, there are also some apparent differences. Most notably, the spectral shape of the newly-formed peak in \src\ is non-thermal, similar to the overall enhancement we observe. By contrast, the SGR~1830$-$0645 phase-averaged and phase-resolved spectra were favorably fit to a thermal model \citep{younes22ApJ:ppm}. This difference can partly be attributed to unfavorable observing geometries of the active loop in SGR~1830$-$0645 which resulted in a weak hard X-ray component and/or the locale of the active region. An equatorial active loop could imply a smaller loop extent, and in turn a fainter hard X-ray component. On the other hand, a transient thermal X-ray component in \src\ could easily be masked by the bright quiescent soft X-ray flux, $L_{\rm X}\gtrsim2\times10^{35}$~erg~s$^{-1}$, in contrast to the much lower quiescent luminosity of SGR~1830$-$0645, $L_{\rm X}<10^{33}$~erg~s$^{-1}$. In short, the spectral differences of the newly-formed peaks between the two sources might reflect simple geometrical and intrinsic discrepancies in the two sources, yet the underlying physical picture of the outburst is rather consistent.

The pulse-peak migration in SGR~1830$-$0645 was interpreted as either plastic motion of the crust, or relaxation of twisted coronal field loops. We consider these two scenarios again in \src\, although one should bear in mind that even if one region -- the interior or the exterior -- is the driver of the observed pulse-peak motion, ultimately the crust and corona are dynamically coupled through the magnetic field lines that cross them both.

A slow plastic failure or `flow' of the crust comes about when stresses build up to beyond the elastic yield stress, meaning the crust can no longer absorb further stress whilst remaining static. Magnetic-field evolution in the crust, for field strengths $\gtrsim 10^{14}\,\mathrm{G}$, naturally leads to the development of such large stresses at the age of a typical magnetar, as a consequence of Hall drift \citep{perna11ApJ,2019MNRAS.486.4130L,dehman20,kojima22}. In the scenario that attributes pulse-peak migration to this plastic crustal motion, the observed motion is simply the plastic velocity of the crust -- subject to the caveats about viewing geometry discussed above -- which for a slow and steady flow is proportional to how far above the yield stress the crust is, and inversely proportional to the viscosity of the crust in this phase \citep{2016ApJ...824L..21L,kojima21}. The former factor might seem to indicate that the plastic velocity should decrease gradually as stress is relieved by the plastic flow, rather than remaining roughly constant as indicated by the linear change in pulse-peak location as a function of phase. However, the interplay between plastic flow and Hall drift is complex, with plastic flow not able to remove stress efficiently; simulations show that plastic flow can be relatively steady, or even increase stresses locally \citep{gourlan21}. It is not sensible to draw strong conclusions from the inferred surface velocity for \src\ being half that of SGR 1830--0645. It is, however, worth remarking that this is not unexpected within the plastic-flow scenario. Slower plastic motion suggests a higher viscosity for the plastic phase, which in turn suggests the failing region is deeper in the crust, since this viscosity likely scales with mass density \citep{2019MNRAS.486.4130L}. Crustal failures at greater depths release more energy \citep{lander15mnras}, which is consistent with previous observations of X-ray bursts from \src\ above $10^{39}\,\mathrm{erg}$ \citep{lin11ApJ:1E1841}, in contrast with the relatively feeble bursts seen from SGR 1830--0645.

In conjunction with such plastic motions, one expects that surface and magnetospheric field lines will be mobile, exhibiting time-varying departures from pure dipolar morphology.  This interconnection was alluded to for the first observed migration of pulse peaks seen in SGR 1830$-$0645 \citep{younes22ApJ:ppm}.  Departures from dipolar fields in magnetars were postulated in \cite{thompson02ApJ:magnetars}, wherein a current-activated magnetosphere adds toroidal components $B_{\phi}$ to generate twisted field morphology, a picture based on solar coronal loop contexts \citep{Wolfson-1995-ApJ}. This activation could well be triggered during outbursts by subsurface energy release associated with glitches, plastic flow or crustal failure.  Currents in twisted magnetospheres are concentrated in restricted zones \citep{beloborodov09ApJ} that assume quasi-toroidal shapes in the axisymmetric, ideal MHD approximation \citep{chen17:mag}.  Charges can then release their energy by bombarding the surface at the field line footpoints in the twist regions and thereby generate X-ray hot spots \citep{Beloborodov-2016-ApJ}.  Prior to this, these leptonic charges can radiate in the magnetosphere, colliding with the hot surface emission via the resonant inverse Compton scattering (RICS) process, and if this signal extends to above a few MeV in energy, it can create $e^+e^-$ pairs that can subsequently emit synchrotron radiation in the $\sim 1-50\,$keV band.  These mechanisms for hard X-ray generation are discussed above, and are addressed extensively for this epoch for \src\ in \citet{stewart24arXiv241216036S}. The pulse peaks observed for \src\ in the \nustar\ data of Figure~\ref{fig:nusprof} very probably correspond to such emission from somewhat confined magnetospheric regions.

Once the activation energy source abates, the radiative dissipation and surface bombardment de-energizes the plasma-loaded magnetosphere, and it is expected \citep{beloborodov07ApJ:magCorona,beloborodov09ApJ} that the twisted fields slowly unwind and the volume of the current loops declines due to Ohmic dissipation.  The timescale for this decay is nominally a year or so \citep{beloborodov09ApJ} for global, axisymmetric twisted configurations, and putatively shorter for more constricted twist volumes.  The fields relax progressively towards the long term field morphology, which may possess some twist and not be purely dipolar in magnetars that exhibit persistent hard X-ray emission, as \src\ does.  During this relaxation, the surface locale of the twist footpoints can shrink and migrate towards the polar regions, nominally the first and last locale of enhanced plasma loading of magnetospheres. The magnetospheric twist zones will similarly migrate and volumetrically shrink, consistent with the observed evolution and merging of pulse peaks in Figures~\ref{fig:nusprof} and~\ref{fig:nicerprof}.  The timescale for this pulse profile evolution is somewhat longer for \src\ than that for SGR 1830$-$0645, suggesting that \src\ possessed the larger twist field volume or energy content (or both) during this burst active phase. Future detailed simulations of non-axisymmetric twist set-ups and temporal evolution is therefore highly motivated.  Such theoretical studies will have the goal of assessing the connections between the pulse peak phase separations, the evolution timescales, and the twist enhancement to the net field energy and geometry.

Finally, it is worth noting that while this is the first reported radiative outburst from \src, it might be due to observational biases. Previous pointed X-ray observations of the source following bursting activity, most prominently in 2010, suffered from a limited signal-to-noise or coverage in the soft and hard X-ray band. For instance, the outburst was followed-up with Swift-XRT providing information on the soft X-ray band, within which we only observe a 20\% flux increase with our \nustar\ monitoring \citep[Table~\ref{tab:specres}, ][]{kumar10ApJ:1841,lin11ApJ:1E1841}. Given the typical short exposures of XRT and its low effective area, such an increase could easily be missed. Moreover, few RXTE observations were acquired in the aftermath of the outburst \citep{dib14ApJ}, and these also suffered from a large background and short integration, which may hinder its ability to observe the pulse profile changes we report. Hence, our results highlight the importance of large throughput, low background broadband instruments such as \nicer\ and \nustar. Future generation satellites such as Strobe-X \citep{ray19arXiv190303035R} and HEX-P \citep{alford24FrASS} would provide data with unprecedented quality even for such a modest outburst. This is critical to increase the sample of observed magnetar outbursts from the younger, brighter member of the population for completeness and enable population-level studies of, e.g., pulse-peak migration that is poised to grow.

\subsection{Radio non-detection}

The absence of any detectable radio emission could be explained by the following reasons: the radio emission is too faint to be detectable, the emission is beamed away from our line of sight or the magnetar is simply not emitting at radio wavelengths. In the first case, the distance and the expected scattering from multi-path propagation could reduce the radio luminosity below the sensitivity of the telescopes used in this work. The distance of \src\ is about a factor of two larger than the radio loud magnetars \xteeigthteen\ \citep{camilo06Natur:1810} and \swifteigthteen\ \citep{esposito2020} but similar to the radio loud magnetars \sixteentwentytwo\   \citep{levin10ApJ} and \seventeenfourtyfive\  \citep{2013Natur.501..391E}. Hence, the radio luminosity must be significantly lower than the known sources or impacted by other effects, such as scattering. We estimate the expected scattering timescale using the NE2001 model \citep{cordes2002} to be around \SI{15}{ms} at \SI{1.4}{GHz} (MeerKAT L-band and Band 1 of the Effelsberg UBB observations), which is $\sim$0.1\% of the pulse period and would not significantly impact the folded profile. Scattering impacts the detectability of faint single pulses as long as the pulse broadening timescale is longer than the intrinsic duration. The scattering at \SI{1.4}{GHz} would reduce the observed fluence of a single pulse with an intrinsic duration of $\sim$\SI{1}{ms} by a factor of a few. 
At higher frequencies the scattering time decreases to $\sim$ \SI{1}{ms} for \SI{>2.8}{GHz} (MeerKAT S-band and Effelsberg Band 2-5 observations) and would only impact very narrow pulses.  Additionally, the radio beam might have a geometrical orientation that does not allow to see into the radio beam, i.e. the radio beam rotates such that the line of sight does not intersect the point at which the radio emission is emitted \citep{pulsarhandbook2012,kaspi17:magnetars}. 

However, the magnetar might also be intrinsically radio quiet. According to the fundamental plane of radio emission in magnetars \citep{rea2012}, \src\ lies in the radio-quiet zone as its X-ray luminosity significantly exceeds the spin-down luminosity. One exception to this observational constraint is \sgrnineteen\ which showed luminous ms-scale radio bursts, and a month-long, faint pulsed radio component albeit having an X-ray luminosity exceeding its spin-down luminosity. Interestingly, these appeared following a spin-down glitch \citep{2023NatAs...7..339Y} event indicating that the state of the magnetosphere may play a crucial role in the ignition of magnetar radio emission \citep{thompson08ApJ:radio}. Despite the many more spin-up glitches observed from magnetars \citep{dib14ApJ}, no subsequent radio activation has been observed. Finally, we note that the searches for radio emission presented in this work were more likely to detect pulsar-like (i.e. spin-down powered) emission than the rare, luminous, ms-scale radio bursts observed from \sgrnineteen\ \citep{Bochenek2020,chime2020}, which have provided evidence for a magnetar origin to fast radio bursts (FRBs). Such FRB-like bursts are exceedingly rare from Galactic magnetars, and therefore unlikely to occur during the short observations presented here, but also $\sim$MJy flux densities may be so large as to saturate the receiving system. 

\section*{Acknowledgments}

This material is based upon work supported by the National Aeronautics and Space Administration under Agreement No. 80GSFC24M0006 issued through the Office of Science. G.Y. acknowledge support through NASA grants 80NSSC21K1997, 80NSSC23K1114, and 80NSSC25K7257, which are partly funding PhD students R.S. and A.v.K., and postdoctoral fellow A.M. M.G.B. thanks NASA for generous support under awards 80NSSC24K0589 and 80NSSC25K7257. W.C.G.H. acknowledges support through grant 80NSSC23K0078 from NASA. M. Ng is a Fonds de Recherche du Quebec – Nature et Technologies (FRQNT) postdoctoral fellow. L.G.S. is a Lise Meitner Group Leader, and together with M.L.B. acknowledge support from the Max Planck Society.  The MeerKAT telescope is operated by the South African Radio Astronomy Observatory, which is a facility of the National Research Foundation, an agency of the Department of Science and Innovation. This work has made use of the ‘MPIfR S-band receiver system’ designed, constructed, and maintained by funding of the MPI für Radioastronomy and the Max Planck Society. Observations used PTUSE for data acquisition, storage, and analysis which was partly funded by the Max-Planck-Institut für Radioastronomie (MPIfR). Based on observations with the 100-m telescope of the MPIfR (Max-Planck-Institut für Radioastronomie) at Effelsberg.  The UBB receiver and the Effelsberg Direct Digitisation (EDD) system are developed and maintained by the Max Planck Institute for Radioastronomy (MPIfR) and are funded by the Max Planck Gesellschaft (MPG). G.Y. thanks Alexander Philippov for the enlightening discussions pertaining to the results presented in this manuscript.

\end{document}